\documentclass[sigconf,natbib=true]{acmart}

\usepackage[utf8]{inputenc}
\usepackage{amsmath}
\usepackage{graphicx}
\usepackage{hyperref}
\usepackage{graphicx} % Required for inserting images
\usepackage{adjustbox} 
\usepackage{booktabs}
\usepackage{multirow}
\usepackage{color}
\usepackage{amsmath}
\usepackage{balance}

\AtBeginDocument{%
  }

\setcopyright{none}
\renewcommand\footnotetextcopyrightpermission[1]{} % removes footnote with conference information in first column
\acmConference[]{arxiv}{May 2026}{version 1}

\begin{document}

\title{Plans for Evaluating Structured Generative Search Summaries}


\author{Tetsuya Sakai}
\email{tetsuyasakai@acm.org}
\affiliation{%
  \institution{Waseda University/Naver Corporation}
  \city{Tokyo}
    \country{Japan}
}

\author{Jina Lee}
\email{jinalee1697@ruri.waseda.jp}
\affiliation{%
  \institution{Waseda University}
  \city{Tokyo}
    \country{Japan}
}

\author{Hanpei Fang}
\email{hanpeifang@ruri.waseda.jp}
\affiliation{%
  \institution{Waseda University}
  \city{Tokyo}
    \country{Japan}
}

\author{Young-In Song}
\email{song.youngin@navercorp.com}
\affiliation{%
  \institution{Naver Corporation}
  \city{Seongnam}
    \country{Korea}
}

\renewcommand{\shortauthors}{Tetsuya Sakai et al.}

\begin{abstract}
We propose a framework
for evaluating structured generative search summaries
that are placed atop organic web search results.
A structured summary, generated by a large language model,
typically consists of an overview, 
several sections with section titles,
and a list of source documents that are cited within the summary.
We then describe our plans for implementing and evaluating the framework.
\end{abstract}



\maketitle

\section{Introduction}\label{s:intro}

\begin{figure}[t]
\begin{center}

\includegraphics[width=0.49\textwidth]{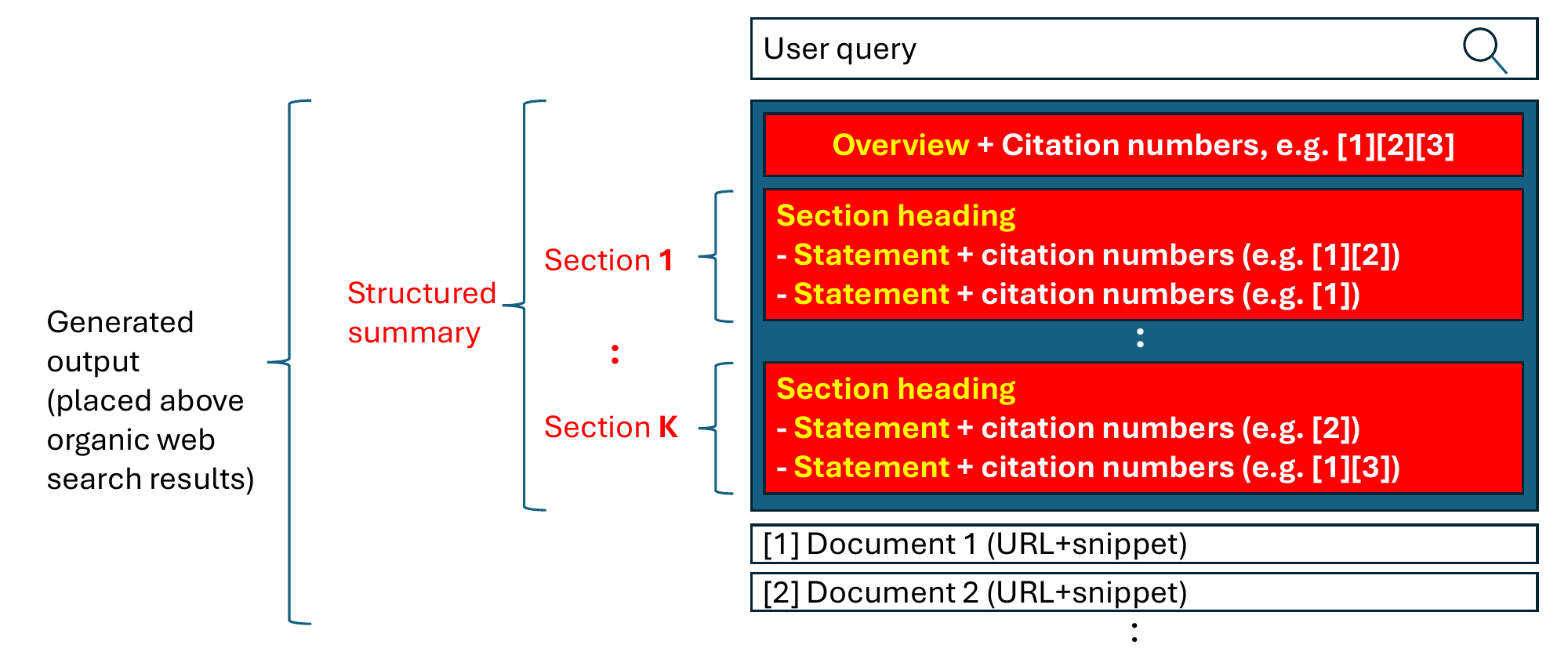} 
\caption{A typical structure of a  generated output included in a 
SERP, with a structured summary containing an overview and one or more sections.
}\label{f:serp}

\end{center}
\end{figure}

The advance of Large Language Models (LLMs) is rapidly changing the experience of web search engine users.
To try to quickly satisfy the user's information need without making them read retrieved web pages,
search engine companies have started to include a \emph{generated output} in the Search Engine Result Page (SERP) for some queries,
atop traditional ``organic'' web search results.
Figure~\ref{f:serp} shows a typical structure of a generated output
that comprises a \emph{structured summary} and a \emph{doclist} (i.e., a list of cited documents).
Unlike \emph{plain-text} search engine outputs that have been considered in earlier studies (e.g.,~\cite{sakai2011cikm}), 
the structured summary starts with a short textual \emph{overview} that is designed to address the user's query directly,
followed by one or more textual \emph{sections}, 
where each section consists of a
\emph{section heading} that briefly describes what the section is about,
and a  \emph{section body} that contains one or more \emph{statements}.
In general,
the overview and each statement
are accompanied by zero or more 
\emph{citation numbers}.
Each citation number points to one of the documents in the doclist
shown beneath the structured summary.
Thus the doclist is a short list of \emph{cited} web pages,
typically representing a subset of 
the traditional organic web search results 
that are placed below it.

For over two decades, search engine companies have used 
nDCG and nDCG-like measures to evaluate and optimise \emph{document lists}: see, for example, \citet{burges2005}.
However, as generated outputs in SERPs like the one illustrated in Figure~\ref{f:serp} are becoming more common than ever,
we need a sound framework to evaluate such outputs.
In light of this situation,
the present study addresses the question of evaluating the structured summary part of a SERP,
i.e., the red parts shown in Figure~\ref{f:serp}, by treating the doclist as a given.
Put another way, our concern is:
\emph{Given this doclist, how good is this structured summary? Which parts of it are good/bad and why? How can we improve it?}
This evaluation problem is particularly important because LLMs are known to ``hallucinate''~\cite{kalai2025}\footnote{
We have no intention of anthropomorphising LLMs~\cite{forster2026}: hence the quotation marks.
} 
even if the source documents are accurate,
misinformation can be injected at the text generation step, which may have a substantial negative impact on the user.
%Moreover, for some queries, evaluating the \emph{correctness} or \emph{relevance} of the returned information may not be sufficient.
%For example,
%if several different (and possibly conflicting) viewpoints are relevant to the query,
%then
%the generated text should try to cover them rather than just providing one viewpoint (\emph{diversity});
%if different entities deserve to be included in the generated text
%in the context of a given query,
%then the generated text should ensure exposure of marginalised entities (\emph{fair exposure}).
That is, not only \emph{relevance} to the information need
but also \emph{faithfulness}~\cite{cao2024,es2024,gao2023,lamsiyah2025,li2025,liu2023,rashkin2023,vandort2026,zhou2023} to the source documents should be considered.
Moreover, for some types of queries,
evaluating the \emph{comprehensiveness} of the structured summary will be needed:
for example, we would like to be able to ask:
\emph{For this query, do these sections of this structured summary cover all aspects that should be covered?}
Evaluating the relevance and faithfulness of parts of the structured summary
is closely related to the traditional notion of \emph{precision}, which requires systems to reduce \emph{noise} in the search results;
Evaluating the comprehensiveness of the entire structured summary
is closely related to the traditional notion of \emph{recall}, which requires systems to prevent \emph{missing} information that should not be missed.

%The contribution of the present study are as follows.
%\begin{itemize}
%\item We provide a general evaluation framework for quantify the quality of structured generated texts such as the one shown in Figure~\ref{f:serp},
%within the context of a given query and a doclist.
%Our framework evaluates the generated text from two angles: a user model-based 
%method that rewards presentation of relevant and faithful information near the beginning of the text (``top heaviness''),
%and a pooling-based method that quantifies the comprehensiveness of the entire generated text.
%\item We demonstrate the usefulness of our evaluation framework by comparing it with baselines in terms of how often the measures agree with humans in %terms of SERP preferences,
%using real Korean SERPs. The comparison with the baselines show the effectiveness of our utilisation of the text structure, our top-heaviness scheme, and %our quantification of comprehensiveness.
%\item As the Korean data set used in our experiments is proprietary,
%we make publicly available a surrogate data set that masks all Korean texts and URLs,
%but equipped with the labels collected in our experiments and the effectiveness scores.
%\end{itemize}

In this paper,
we propose a framework
for evaluating structured generative search summaries
that are placed atop organic web search results,
like the one shown in Figure~\ref{f:serp}.
%A structured generative summary, generated by a large language model,
%typically consists of an overview, 
%several sections with section titles,
%and a list of source documents that are cited within the summary.
We then describe our plans for implementing and evaluating the framework.
The outcomes of our implementation and evaluation will be reported elsewhere.

\section{Related Work}\label{s:related}

\subsection{Summarisation and Complex Question Answering Evaluation}

%Historically, we had \emph{extractive} summaries (e.g.. those based on \emph{sentence selection})
%and \emph{abstractive} ones, i.e., those that rely in some way on paraphrasing or text generation: see, for example,~\citet{hovy1997}.
%Clearly, summaries generated by LLMs fall into the latter, more advanced category.

Historically, text summarisation and question answering tasks
had a few aspects in common regarding how the systems' textual responses were evaluated.
Firstly, it was assumed that the \emph{ground truth} data (i.e., manually-composed reference summaries or 
manually-identified correct answers)
were available. Second, the comparison
of the system response and the ground truth
were performed by decomposing both into smaller units
such as N-grams~\cite{lin2004}, \emph{semantic content units}~\cite{nenkova2007}, or \emph{nuggets}~\cite{hoatrangdang2007,mitamura2010,rajput2011,voorhees2004}.
Among the nugget-based approaches,
our work is inspired by the S-measure/U-measure frameworks~\cite{sakai2011cikm,sakai2013sigir},
which are unique in that they take into account the \emph{position} of each nugget within the summary for the evaluation.

The modern challenge in evaluating system-generated texts, however,
 is that we often do not have access to the ground truth; it may not even exist.
 This is also the case in our study,
 where the task is to evaluate generated summaries 
 for \emph{any} web search query entered by the users.
 Thus we need a \emph{reference-free}~\cite{es2024,joko2026} evaluation approach.

\subsection{RAG and Conversational Search Evaluation}\label{ss:RAGevaluation}

RAG (Retrieval-Augmented Generation) combines the strengths of IR 
and the generation capability of LLMs~\cite{lewis2020,petroni2024}.
As LLMs can now extract nuggets from text with reasonable accuracy,
\citet{pradeep2025} ``refactored'' the nugget-based evaluation method from the TREC 2003 question answering track~\cite{lin2004}
for the TREC 2024 RAG track; see also \citet{alaofi2024,lajewska2025}.
Nugget-based and similar RAG evaluation methods were also adopted 
at the TREC iKAT 2024 track~\cite{abbasiantaeb2025} and the TREC 2024 NeuCLIR track~\cite{samarinas2025}.
Other nugget-based 
RAG evaluation efforts include the ongoing NTCIR-19 R2C2 task~\cite{sakai2025brev-rag}.

%\subsection{Conversational Search Evaluation}

\emph{Conversational search}~\cite{radlinski2017}
(or \emph{Conversational Information Access}~\cite{joko2026})
generally 
 involves \emph{multi-turn} system responses.
 (See also pre-LLM-era efforts in \emph{interactive question answering}~\cite{hoatrangdang2007}.)
The present study focusses on \emph{single-turn} responses,
i.e., a single generated summary for a given query.
However, our work builds on a user model proposed
in the context of conversational search, namely,
that described in
\citet{sakai2025evia}.
In their conversational search user model,
a population of users with the same information need
examines a ranked list of \emph{nuggets}, which represent parts of the system responses (which can be multi-turn), from top to bottom.
Some of the users abandon the list after seeing a particular relevant nugget; others continue on.\footnote{
The user model is similar to that of the Sakai/Robertson \emph{Normalised Cumulative Utility} (NCU) measures~\cite{sakai2008evia};
the main difference is that the NCU user model is for ranked list of \emph{documents} such as web pages.
}
Instead of using nuggets as the basic evaluation unit,
we leverage the structured nature of our generated summaries,
and use ``line numbers'' in the summary as the basic unit, as we shall describe in Section~\ref{ss:XUX}.
Inspired by the aforementioned S-measure/U-measure approach,
we use line numbers as the positions of relevant pieces of information within the summary.

\subsection{Evaluation Axes}\label{ss:axes}

The SWAN (Schematised Weighted Average Nugget) framework~\cite{sakai2023swan}
lists 21 evaluation criteria for conversational systems,
including \emph{groundedness} (i.e., Is the claim supported by a piece of evidence?)~\cite{menick2022},
\emph{sufficiency} (i.e., Does the system's turn satisfy the requests in the previous user turn?)
\emph{modesty} (i.e., Is the system's answer confidence score appropriate?)~\cite{sakai2024emtcir,sakai2025brev-rag},
\emph{fair exposure} (i.e., Does the system mention entities from different groups fairly in its responses?)~\cite{sakai2025evia},
and
\emph{harmlessness} (i.e., ``\textit{no stereotypes, no microaggressions, no threats, no sexual aggression, no insults, no hate or harassment}''~\cite{glaese2022}).
See also the HHH (helpful, honest, and harmless) criteria of \citet{askell2021}.

\citet{joko2026} take an approach related to SWAN for evaluating conversational information access,
where not only turn-level
axes such as \emph{interestingness} (i.e., \emph{engagingness} of SWAN)
but also dialogue-level axes such as \emph{task completion} and \emph{overall impression}
are considered.
The Dialogue Quality subtasks of the NTCIR DialEval tasks~\cite{tao2022,zeng2020}
also used related criteria, including \emph{task accomplishment} and \emph{customer satisfaction},
in the context of evaluating helpdesk-customer (human-human) dialogues.

In terms of evaluation criteria,
our framework concerns
\emph{relevance} to the user's turn,
\emph{faithfulness}~\cite{cao2024,es2024,gao2023,lamsiyah2025,li2025,liu2023,rashkin2023,vandort2026,zhou2023} with respect to the cited documents
(which we equate with groundedness),\footnote{
\citet{rashkin2023} define a related concept called \emph{attribution}.
Also, while \citet{wallat2025} define \emph{citation faithfulness}
based on whether the model actually relies on cited documents rather than post-rationalising the citations,
this is out of the scope of this present study as our evaluation concerns how the summaries look like from the user's viewpoint.
}
and 
\emph{comprehensiveness} which is similar to \emph{sufficiency}~\cite{sakai2023swan} and \emph{coverage}~\cite{gienapp2024}.
While it is relatively easy to discuss the \emph{relevance} and \emph{faithfulness}
of a given summary,
that is not the case with \emph{comprehensiveness},
since we generally lack the ground truth and therefore there is no notion of \emph{recall base}.
Hence, to define comprehensiveness,
we require \emph{multiple} summaries per user query, 
so that a concept akin to \emph{relative recall}~\cite{clarke1997} can be leveraged.
We shall define comprehensiveness in Section~\ref{ss:comprehensiveness}.

\subsection{Reading Models}

The SIGIR 2024 perspectives paper by \citet{gienapp2024} has a movitation similar to ours 
except that they do not discuss the structured nature of generated summaries for a given query.
They propose a user \emph{reading model} which is a
``\textit{monotonically decreasing weight function over statements that discounts the contribution of later statements in a response}''
to reward
``\textit{putting the most important pieces of information first.}''
This is exactly what the aforementioned S-measure from 2011~\cite{sakai2011cikm} featured.
In this paper,
we also employ a user reading model,
but ours is based on the assumption that there is a population of users reading a \emph{structured} summary,
as mentioned in Section~\ref{ss:RAGevaluation}.

\subsection{LLMs Evaluating LLMs}

There is a controversy regarding the evaluation of LLMs \emph{using} LLMs~\cite{clarke2025evia,faggioli2023,panickssery2024,sakai2025cikm,soboroff2025,upadhyay2024initial}.
In the context of web search,
clearly our goal is to satisfy the \emph{user}, not an LLM.
However, our view is that LLMs can be leveraged effectively when used within a
\emph{divide and conquer} evaluation approach:
``\textit{if LLMs’ behaviours are not altogether predictable, then we should try to
reduce the number of confounding uncertainty factors, by giving them simple and controllable subtasks
instead of a complex task}''~\cite{sakai2025brev-rag}.
Thus,
whenever we require \emph{labellers},
we hire either humans or LLMs 
while ensuring that the labelling tasks are relatively straightforward.

\section{Proposed Evaluation Framework}\label{s:proposed}

Unlike existing studies that explore the evaluation of ``flat'' text with citations (e.g.~\cite{gienapp2024}),
our evaluation framework evaluates the \emph{structured summary} part of a SERP 
in the context of a given query and a doclist, as illustrated in Figure~\ref{f:serp}.
The structured summary generally consists of an \emph{overview} $o$
followed by $I (\geq 0)$ \emph{sections},
where each section $\textit{sec}_{i}$ ($i = 1, \ldots,  I$)
contains a \emph{section heading} $h_{i}$
and $J_{i} (>0)$ \emph{statements} $s_{ij} (j=1, \ldots, J_{i})$.
The overview and the statements are accompanied by zero or more citation numbers.
(Instead of an overview at the beginning of the structured summary,
a short ``summary'' may be placed at the end: our evaluation framework accommodates both.
Hereafter, we shall focus on structured summaries with an initial overview.)

Sections~\ref{ss:XUX}-\ref{ss:top-heavy}
describe our \emph{Expected User Experience} (XUX) measure,
which evaluates a given structured summary from a user perspective.
However, as XUX cannot quantify neither ``relevant information that is missing in the text''
nor ``balanced coverage of multiple aspects relevant to the information need,''
Section~\ref{ss:comprehensiveness}
describes our \emph{Comprehensiveness} (Comp) measure
which is designed to address these questions by
``pooling'' sections from multiple structured summaries;
this is similar in spirit to estimating the recall of multiple 
``10-blue-links''
search engines~\cite{clarke1997}
and to the traditional document pooling for constructing IR test collections~\cite{sparckjones1975,sakai2019clef,voorhees2005chapter2}.
Finally, Section~\ref{ss:SGSS} defines 
our quick summary measure, SGSS (Structured Generative Search Summary) score,
as a combination of XUX and Comp.

\begin{figure}[t]
\begin{center}

\includegraphics[width=0.49\textwidth]{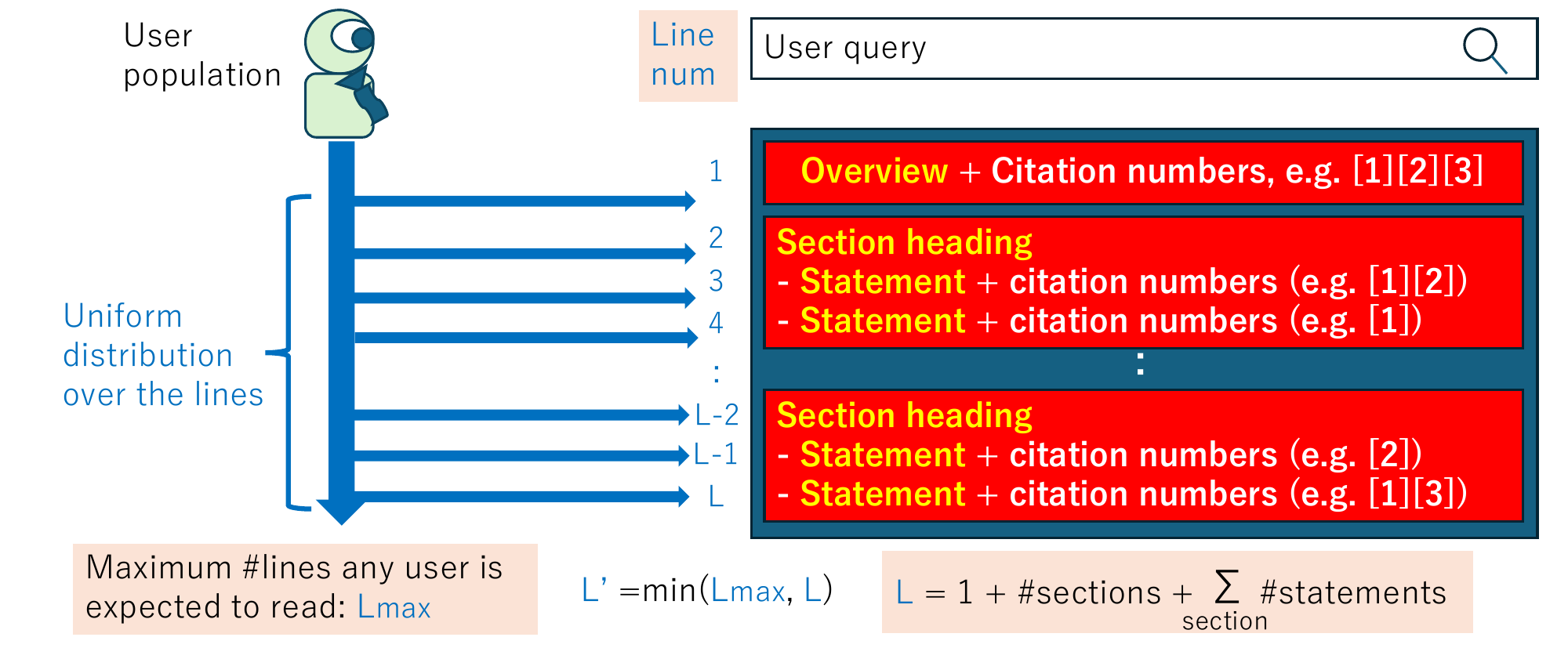} 
\caption{The user model of our SGSS evaluation framework.
}\label{f:usermodel}

\end{center}
\end{figure}

%\section{Overview of the Framework}\label{ss:overview}

\subsection{Expected User Experience}\label{ss:XUX}

For a given query and a structured summary $S$,
we envision a \emph{population} of users each having an information need represented by that query.
We assume that users generally read $S$ starting from the top,
and that different users abandon (i.e., give up reading) $S$ at different \emph{positions} within $S$.\footnote{
This \emph{linear} reading behaviour assumption is analogous to the \emph{linear traversal} assumption
behind traditional IR evaluation measures.
}
To define an \emph{Abandonment Probability Distribution} (APD) for the user population over different positions within $S$,
let us first define \emph{line numbers} within $S$, where
the overview ($o$), section headings ($h_{i}$'s), and statements ($s_{ij}$'s) each constitute one line.
Then, 
%for a generated text containing an overview and $I$ sections,
%where the $i$-th section ($i=1,\ldots, I$) contains $J_{i}$ statements,
the total number of lines in $S$ is given by
\begin{equation}\label{eq:L}
L = 1 + I + \sum_{i=1}^{I} J_{i} \ ,
\end{equation}
as depicted in the right part of Figure~\ref{f:usermodel}.

In ranked list evaluation in IR, we often apply a document cutoff $d$ to SERPs and consider only the top $d$ documents.
Similarly, let us introduce $L_{\mathrm{max}}$, the maximum number of lines in a structured summary
any user is expected to read. In effect, this truncates the structured summary to only top $L_{\mathrm{max}}$ lines 
in order to
penalise excessively long texts. (If the truncation is unnecessary, let $L_{\mathrm{max}}=\infty$.)
%; later,
%we instantiate $L_{\mathrm{max}}$ based on a search engine user log.)
Let $L'=\min(L, L_{\mathrm{max}})$.
As a simple instantiation of the APD,
we consider a \emph{uniform} distribution over $L'$, as depicted in the left part of Figure~\ref{f:usermodel}.
That is, we assume that the users are equally likely to stop reading at any line at or above Line $L'$,
due to satisfaction or dissatisfaction or for some other reason.
This generalises the uniform distribution over \emph{relevant} documents
used in Average Precision and Q, as described in the aforementioned Sakai/Robertson NCU suite of IR measures~\cite{sakai2008evia}.

One way to determine $L_{\mathrm{max}}$ would be to 
observe real user behaviour data over the summaries,
such as clicks, scrolls, and hovers with timestamps.
Another approach would be to follow the S-measure approach~\cite{sakai2011cikm},
where it is assumed that the users' reading speed is constant, and that 
the user has a $M$-minute window to satisfy their need.
For example, from \citet[p.206]{rhiu2017},
we can assume that the typical reading speed for Korean texts is about 500 characters per minute;
hence we can assume that the user will read up to $500M$ characters;
if the average character length per line in a Korean summary is $\textit{avlen}$,
we can let $L_{\mathrm{max}} \approx 500M/\textit{avlen}$.

Having defined the probability that users will abandon the structured summary at Line $l$,
let us now consider the quality of the summary from the viewpoint of these different users.
More specifically,
for the group of users who abandon the summary after reading Line $l$,
let $X(l)$ denote their overall user experience, which we shall instantiate in the following sections.
Then, we can compute \emph{Expected User Experience} (XUX)
as an expectation of $X(l)$ over the aforementioned uniform probability distribution.
\begin{equation}\label{eq:XUX}
\textit{XUX}  = \frac{1}{L'} \sum_{l=1}^{L'} X(l) \ .
\end{equation}

We can also consider a simplified variant of XUX with a modified APD,
by assuming that users will abandon the structured summary only after reading the overview line 
or after reading an {entire section}.
Let $F(l)$ be one if line $l$ is either an overview line or the Final line of a section, and zero otherwise.
Again, by assuming a uniform probability mass function over the lines where $F(l)=1$,
we can define:
\begin{equation}\label{eq:XUX-F}
\textit{XUX}_{F} = \frac{1}{1+I}  \sum_{l=1}^{L'} F(I) X(l) \ .
\end{equation}
Note that, for simplicity, XUX$_{F}$ does not consider  $L_{\mathrm{max}}$.

Clearly, XUX score can also be Generalised so that 
we can employ any valid APD over the line numbers:
\begin{equation}\label{eq:XUXgeneral}
\textit{XUX}_{G} = \sum_{l} \textit{Pr}(l) X(l) \ ,
\end{equation}
where $\textit{Pr}(l)$ is the probability that the users of a given query will
abandon the summary at Line~$l$, such that $\sum_{l=1}^{L'}\textit{Pr}(l)=1$.
Considering  non-uniform APDs for $\textit{Pr}(l)$ will be left
for future work; however, our view is that evaluation measures should be as simple and interpretable as possible.

%The uniform probability mass functions described above 
%implicitly encourage systems to provide useful information 
%as ``quickly'' as possible to the user, in terms of \emph{where} in the generated text the information should be presented.
%This is analogous to the top-heaviness of Average Precision which also employs a uniform user distribution~\cite{robertson2008}.

%In Section~\ref{ss:user-experience},
%we shall describe how we compute $X(l)$ from 
%labels collected for 
%the overview, section headings, and the statements of the generated text.
%Hereafter, a ``labeller'' means either an LLM or a human assessor.

\subsection{User Experience Building Blocks}\label{ss:user-experience}

Given a single structured summary $S$ like the one shown in Figure~\ref{f:serp}, 
we quantify the user experience $X(l)$ for each line $l$ of $S$ as described below.
%Recall that, as illustrated in Figure~\ref{f:serp},
%a generated text consists of an \emph{overview}
%and one or more \emph{sections},
%with each section consisting of one or more \emph{statements}.
Hereafter, we shall refer to \emph{labellers} for defining the building blocks for $X(l)$:
they can be human, an LLM, or a combination of the two.

\subsubsection{Overview Quality}\label{sss:OQ}

The overview part of the structured summary is generally short,
and is accompanied by zero or more citations.
We let labellers rate the overview quality using the following criteria, 
where each criterion is on a 3-point scale (Perfectly/Partially/No).
\begin{description}
\item[Overview Sufficiency (OS)] \textit{Does the overview satisfy the information need behind the query?}
\item[Overview Faithfulness (OF)] \textit{Do the cited documents (when taken together) \emph{entail} the overview's claims?}
\item[Overview Representativeness (OR)] \textit{Does the overview actually serve as a good overview of the sections?}
%\item[Coherence with Sections]
\end{description}
Thus, OS, OF, and OR are about
the query-overview relationship, 
the overview-doclist relationship,
and the overview-sections relationship, respectively.
We then map the labels to scores in the 0-1 range.

For overview $o$,
let $\textit{OS}(o), \textit{OF}(o), \textit{OR}(o)$ denote their OS, OF, and OR scores.
The overview quality of $o$ can be expressed as:
\begin{equation}\label{eq:OQ}
\textit{OQ}(o) = w_{\mathrm{OS}} \textit{OS}(o) + w_{\mathrm{OF}} \textit{OF}(o) + w_{\mathrm{OR}} \textit{OR}(o) \ , 
\end{equation}
where $w_{\mathrm{OS}}, w_{\mathrm{OF}}, w_{\mathrm{OR}}$ are weights, which we shall discuss later.

Note that \emph{relevance} is a necessary but not sufficient condition for \emph{sufficiency}:
we ask whether the overview \emph{satisfies} the information need.
As described in Section~\ref{sss:SQ}, we quantify \emph{relevance} at the statement level.

\subsubsection{Heading Quality}\label{sss:HQ}

As each section contains one section heading and one or more statements,
we quantify the quality of each heading in terms of 
\textbf{Heading Representativeness (HR)} as follows.
For the $j$-th statement  ($j \leq J_{i}$) of the $i$-th section ($i \leq I$),
we obtain a relevance grade $\textit{rel}(h_{i}, s_{ij})$ by asking the labellers
\textit{how relevant the statement is to the section heading} on a 3-point scale (Perfectly/Partially/Not relevant).
We then map the labels to graded relevance values $g(h_{i}, s_{ij})$ in the 0-1 range
and finally compute the average relevance of the statements to the section heading.
\begin{equation}\label{eq:HRep}
\textit{HR}(h_{i})=\frac{\sum_{j}g(h_{i}, s_{ij})}{J_{i}} . \
\end{equation}
Thus, HR is about the heading-statement relationship.

\subsubsection{Statement Quality}\label{sss:SQ}

We label each statement in each section using the following criteria, where each criterion is on a 3-point scale (Perfectly/Partially/No).
\begin{description}
\item[Statement Relevance (SRel)] \textit{Is the statement relevant to the information need behind the query?}
\item[Statement Faithfulness (SF)] \textit{Do the cited documents (when taken together) \emph{entail} the statement?}
\end{description}
Thus, SRel is about the query-statement relationship,
while SF is about the statement-doclist relationship.
We then map the labels to scores in the 0-1 range; for statement $s_{ij}$.
We denote the SRel and SR scores of statement $s_{ij}$ by  $\textit{SRel}(s_{ij}), \textit{SF}(s_{ij})$.

%NOTE:
%To penalise nonrelevant and unfaithful statements, we should consider assigning \emph{negative} values to them.

%\subsubsection{Generated Text Conciseness}
% difficult to implement as we don't have minimal text lengths for the nuggets.

\subsection{Top-heavy XUX Instantiations}\label{ss:top-heavy}

Let $c$ be a \emph{component} of a given structured summary,
which can be an overview, a section heading, or a statement.
Let $\textit{line}(c)$ denote the line number of component $c$.
We instantiate the XUX formula (Eq.~\ref{eq:XUX}) with 
\begin{equation}\label{eq:Xinstantiation}
X(l) = \frac{X'(l)}{l} \ ,
\end{equation}
\begin{displaymath}
X'(l)= 
\textit{OQ}(o) 
+ w_{\mathrm{HR}} \sum_{i, \, \textit{line}(h_{i}) \leq l} \textit{HR}(h_{i})
\end{displaymath}
\begin{equation}\label{eq:xprime}
+ w_{\mathrm{SRel}} \sum_{i,j, \  \textit{line}(s_{ij}) \leq l} \textit{SRel}(s_{ij})
+ w_{\mathrm{SF}} \sum_{i,j, \  \textit{line}(s_{ij}) \leq l} \textit{SF}(s_{ij})  \ ,
\end{equation}
where the $w_{\bullet}$'s are weights associated with each component.\footnote{
Eqs.~\ref{eq:OQ} and~\ref{eq:xprime}, when taken together,
do 
not guarantee $X(l) \leq l$
unless 
$w_{\mathrm{OS}} + w_{\mathrm{OF}} + w_{\mathrm{OR}} \leq 1$ and
$w_{\mathrm{HR}}  \leq 1$ and
$w_{\mathrm{SRel}} + w_{\mathrm{SF}} \leq 1$,
so that each line contributes a total score no larger than 1 to $X'(l)$.
Hence, without these constraints,
XUX (Eq.~\ref{eq:XUX}) may potentially attain a value higher than 1.
We do not impose these constraints unless a strictly normalised measure is necessary.
}
%which we shall determine based on regression.
We shall briefly describe one method for setting these weights in 
Section~\ref{s:plans}.

\begin{figure}[t]
\begin{center}

\includegraphics[width=0.49\textwidth]{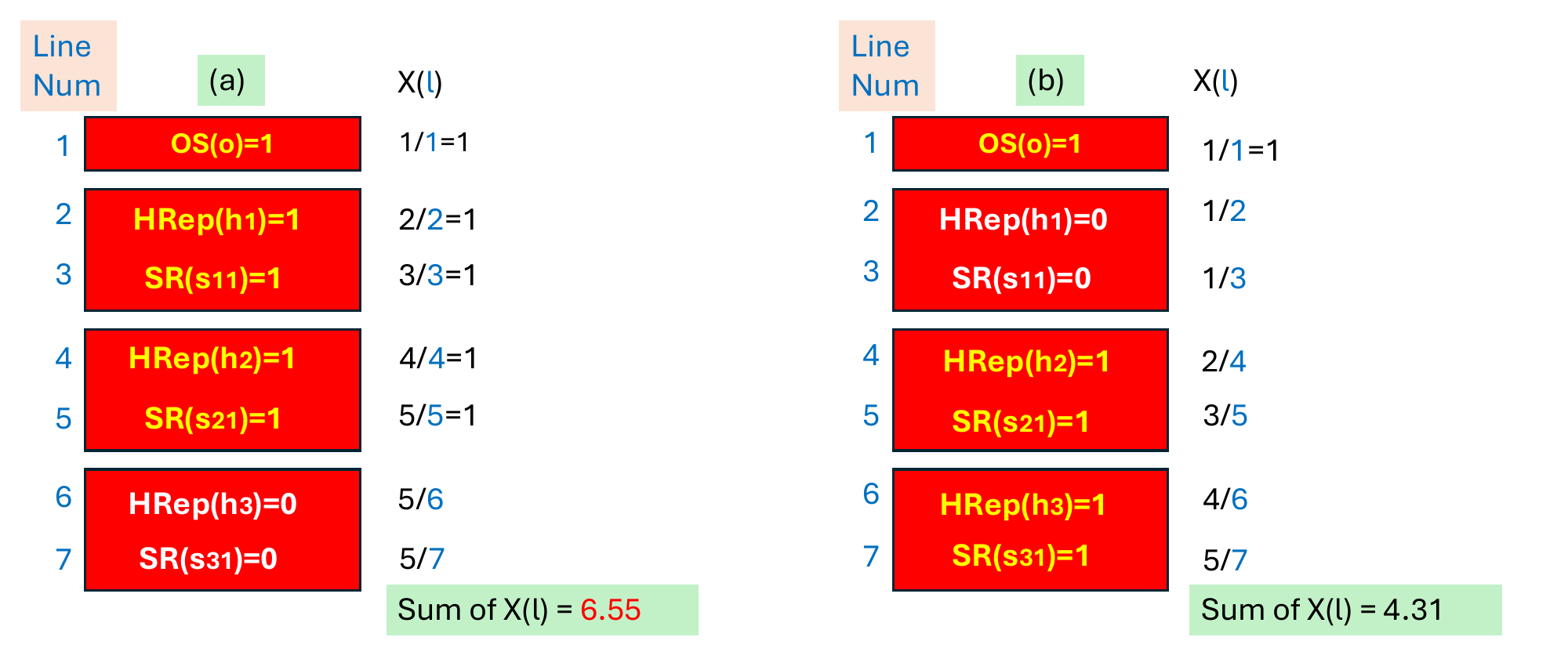} 
\caption{The top heaviness of XUX: Structured summary (a) will receive a higher score than (b).
}\label{f:topheaviness}

\end{center}
\end{figure}

It is important to note that Eq.~\ref{eq:Xinstantiation} 
makes XUX top-heavy, just as the use of \emph{precision}  makes \emph{average precision} top heavy~\cite{robertson2008}
(and the use of the \emph{blended ratio} makes \emph{Q-measure} top heavy~\cite{sakai2014promise}). 
That is, this instantiation rewards \emph{early presentation of good content}.
To see why,
let us take the structured summaries illustrated in Figure~\ref{f:topheaviness} as examples,  where, for simplicity,
each text contains three sections, each with only one statement,
and $w_{\mathrm{OS}}=w_{\mathrm{HRep}}=w_{\mathrm{SR}}=1, w_{\mathrm{OF}}=w_{\mathrm{OR}}=w_{\mathrm{SF}}=0$.
That is, we ignore OF, OR, and SF scores for now.
%That is, we ignore the faithfulness scores for now.
Furthermore, suppose that the OS, HRep and SR scores are simply either 1 or 0 as indicated in the figure.
The only essential difference between texts (a) and (b) is 
the ordering of the three sections,
where exactly one of these sections was given zero scores.
The figure shows how user experience $X(l)$ is calcuated at each line $l$;
it can be observed that the sum of $X(l)$ is higher for (a) than that for (b).
The top heaviness is achieved by 
(i)~the denominator $l$ of $X(l)$, which discounts the user experience 
if a good content moves to a later line in the text; and
(ii)~the numerator $X'(l)$ (Eq.~\ref{eq:xprime}) which ensures
that a good content at Line $l$ contributes also to Line $(l+1)$ and below.

\subsection{Comprehensiveness}\label{ss:comprehensiveness}

In 1973, Cooper, in the context of discussing the choice of IR evaluation measures,
remarked: ``\textit{Why should the status of documents which the system user has not examined have any influence on a user-oriented performance rating?}''~\cite[p.95]{cooper1973}. (See also \citet{zobel2009} from 2009.)
The XUX measure discussed above aligns with the above ``precision-oriented'' perspective
since it only evaluates a given structured summary, i.e., what the user actually sees.
However, from a system developer (e.g., search engine company) point of view,
we would also like to ask ``\textit{What is missing in the structured summary? What else could we have offered to the user?}'': a ``recall-oriented'' perspective.

As we have discussed in Section~\ref{ss:axes},
we are dealing with reference-free evaluation and therefore 
we have to establish a notion similar to \emph{relative recall}.
In order to do this, we require that we have \emph{multiple} summaries for a given query
so that we can ``pool'' information contributed by each of them.
More specifically, as we are evaluating structured summaries,
we pool \emph{sections} from multiple structured summaries to compare them in terms of \emph{comprehensiveness}
as described below.
As each section can be regarded as a representation of an \emph{aspect} or \emph{perspective} relevant to the information need,
comprehensiveness evaluation asks ``\textit{how many of available relevant aspects are covered by this particular structured summary?}''
Here, note that a section from Summary~$a$ and one from Summary~$b$
may not be identical and yet represent the same aspect;
a section from Summary~$a$ may even be roughly equivalent to multiple sections from Summary~$b$.
Thus, the situation is much more complicated 
than traditional document pooling that merely involves merging document IDs
or the task of identifying \emph{near-duplicate passages} (or passage \emph{equivalence classes})~\cite{craswell2024}.
Our approach is designed to address these issues.

%In traditional summarisation evaluation that relied on \emph{reference summaries} (i.e., gold summaries),
%it was possible to define recall-oriented measures (e.g., ``\textit{How many N-grams from the gold summaries do the system summary cover?}~\cite{lin2004}).
%As our evaluation task lacks the notion of gold data, recall-oriented evaluation is not straightforward.
%However, if \emph{multiple} structured summaries are available for a given query,
%we can quantify the \emph{comprehensiveness} of the structured summary 
%in a way similar to how a recall base is defined based on \emph{pooled documents}~\cite{sparckjones1975,voorhees2005chapter2}.
%Comprehensiveness is similar to \emph{diversity} (i.e., whether the text covers different user intents behind the same query)~\cite{clarke2013,sakai2011sigir}
%and \emph{group fairness} (i.e., whether the text provides appropriate exposure to different entities relevant to the query)~\cite{ekstrand2022,sakai2023gfr};
%we use the word comprehensiveness to mean
%how much of different ``aspects'' relevant to the information need are actually covered by means of sections within the structured summary,
%as well as how ``fairly'' they are covered, as we shall elaborate below.

\begin{figure}[t]
\begin{center}

\includegraphics[width=0.49\textwidth]{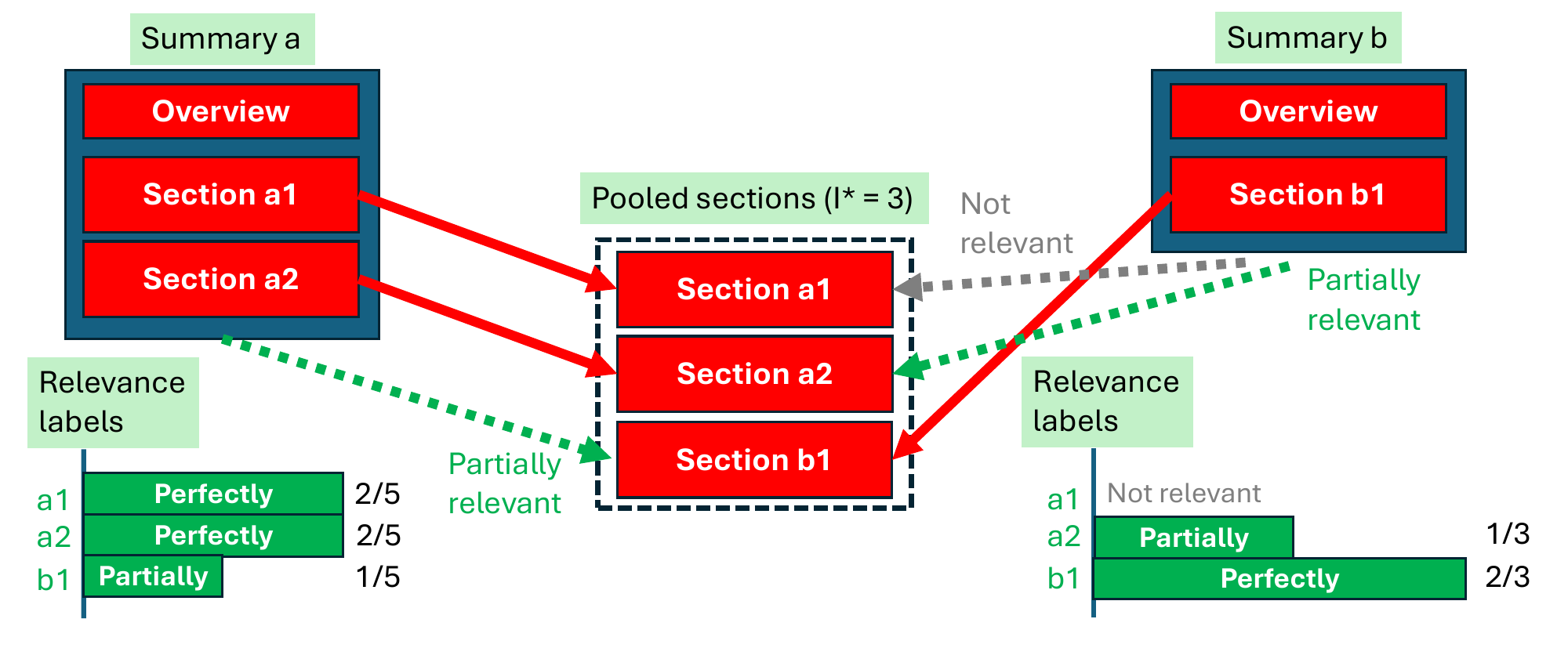} 
\caption{How a pool of section is formed and relevance labels for each pooled sections are obtained.
}\label{f:sectionpooling}

\end{center}
\end{figure}

Given $N (>1)$ structured summaries $S^{1}, \ldots, S^{N}$ for the same query $q$,
we first pool the sections.
Let $\textit{sec}^{n}_{i}$ denote the $i$-th section of Summary $S^{n}$
which contains $I^{n}$ sections ($i=1, \ldots, I^{n}$).
For comprehensiveness evaluation, we require that each summary contains at least one section ($I^{n}>1$).
The pool $P$ for $q$  is defined as 
\begin{equation}
P = \bigcup_{n=1}^{N} \bigcup_{i=1}^{I^{n}} \{\textit{sec}^{n}_{i}\} \ .
\end{equation}
For example,
in Figure~\ref{f:sectionpooling} (where $N=2$),
Summary $a$ contributes sections $a_{1}$ and $a_{2}$ to the pool,
while Summary $b$ contributes section $b_{1}$ to the pool.

Let $K=|P|$, 
and let $\textit{sec}^{\ast}_{k}$ denote the $k$-th pooled section in $P$ ($k=1, \ldots, K$).
Note that $K \geq N$ as we assume that each summary contributes at least one section.
For each summary $S^{n}$ and each $\textit{sec}^{\ast}_{k}$,
we define a 3-point relevance label (Perfectly/Partially/Not relevant)
$\textit{rel}(S^{n}, \textit{sec}^{\ast}_{k})$
as follows.
\begin{tabbing}
\textbf{If} $\textit{sec}^{\ast}_{k}$ is a section from $S^{n}$
\textbf{then} $\textit{rel}(S^{n}, \textit{sec}^{\ast}_{k})=\textit{Perfectly relevant}$\\
\textbf{else} \= $\textit{rel}(S^{n}, \textit{sec}^{\ast}_{k})=$ label obtained by asking labellers \textit{how}\\
			\> \textit{relevant $S^{n}$ is to $\textit{sec}^{\ast}_{k}$} (even though $\textit{sec}^{\ast}_{k}$ is not a section of $S^{n}$).\\
\end{tabbing}
For example, in Figure~\ref{f:sectionpooling}  (where $K=3$),
Summary $a$ is labelled as partially relevant to Section $b_{1}$;
on the other hand,
Summary $b$ is labelled as not relevant to Section $a_{1}$ but partially relevant to $a_{2}$.
In this way, each \emph{summary} is compared with each \emph{section} from a different summary.

Next, we map the labels $\textit{rel}(S^{n}, \textit{sec}^{\ast}_{k})$ to graded relevance scores $g(S^{n}, \textit{sec}^{\ast}_{k})$,
and further compute:
\begin{equation}
g'(S^{n}, \textit{sec}^{\ast}_{k=1}) = 
\frac{g(S^{n}, \textit{sec}^{\ast}_{k})}
{\sum_{k=1}^{K} g(S^{n}, \textit{sec}^{\ast}_{k})}
\end{equation}
so that 
$\sum_{k=1}^{K} g'(S^{n}, \textit{sec}^{\ast}_{k})=1$.
Thus we obtain a probability mass function $F^{n}$
over  the pooled sections.
Examples are given at the bottom of Figure~\ref{f:sectionpooling}.

Let us assume that the pooled sections are equally important
(just as the TREC Web Tracks (2009-2012) assumed that
the subtopics of a given topic are equally important for search result diversification~\cite{clarke2013}).
Accordingly, we employ a uniform distribution $U$ over the pooled sections as our gold (i.e., target) distribution,
with probability $1/K$ for each pooled section.
The comprehensiveness of $S^{n}$ can then be defined as
\begin{equation}
\textit{Comp}(S^{n}) = 1- \textit{JSD}(F^{n}, U) \ ,
\end{equation}
where $\textit{JSD}(\bullet, \bullet)$ denotes 
the \emph{Jensen-Shannon Divergence} between two probability mass functions~\cite{ekstrand2022,sakai2023gfr}.
Note that Comp measures the \emph{similarity} between $F^{n}$ and $U$.
For example, consider a simple situation where there are only two pooled sections,
each representing two contrasting viewpoints for a particular topic addressed in the query.
The above Comp score will  reward a structured summary that covers both viewpoints in a \emph{balanced} manner.
For the examples in Figure~\ref{f:sectionpooling},
the Comp scores for $a$ and $b$ would be 0.983 and 0.792, respectively,
and therefore $a$ is considered more comprehensive.

\subsection{SGSS}\label{ss:SGSS}

To compare multiple structured summaries for a given query,
plotting XUX scores against the corresponding Comp scores may be useful.
In addition, as a quick summary measure, the two can be combined
to form an overall
Structured Generated Search Summary (SGSS) score.
\begin{equation}\label{eq:combined}
\textit{SGSS}(S) = \textit{XUX}(S) + w_{\mathrm{Comp}} \textit{Comp}(S) \ ,
\end{equation}
where $w_{\mathrm{comp}}$ is a weight that determines the impact of Comp
relative to the weights defined in Eq.~\ref{eq:xprime}.
%determined through regression,
%together with the weights defined in Section~\ref{ss:top-heavy}.
%Although the above measure provides a quick summary,
%we stress that detailed diagnosis based on component measures is more important,
%since it is the latter that will suggest actionable improvements to search engine companies.
%We plan to build a visual interface that provides an overview of the component measures discussed above,
%but this is beyond the scope of the present study.

%\begin{table}[t]
%\begin{small}

%\begin{center}
%\caption{Mean AFARs in percentages and the number of trials where AFAR$=0$ (out of $B=100$).}\label{t:AFARs}
%\begin{tabular}{c|c|r}
%\toprule
%MCP method		&(I) Mean AFAR ($\downarrow$)	&(II) \#zero AFARs ($\uparrow$)\\	
%\hline    
%\textbf{NoMCP}	&6.120			&0\\
%\textbf{T}		&0.030			&84\\
%\textbf{T-M}		&0.009			&94\\
%\textbf{T-MC}	&0.002			&98\\
%\textbf{H}		&0.008			&96\\
%\textbf{H-M}		&0.005			&99\\
%\textbf{H-MC}	&0.002			&99\\
%\textbf{F}		&2.680			&0\\
%\textbf{F-M}		&2.660			&0\\
%\textbf{F-MC}	&2.661			&0\\
%\bottomrule        
%\end{tabular}
%\end{center}
%\end{small}
%\end{table}

%\section{Data}\label{s:data}

%\section{Evaluating the Proposed Framework}\label{s:evaluating}

\section{Plans}\label{s:plans}

We are currently preparing data for implementing and evaluating the SGSS evaluation framework described above.
Below, we describe our ongoing work and near-future plans.

\subsection{Dataset Construction}

We are collecting generated output data (See Figure~\ref{f:serp}) for real search engine queries.
As comprehensiveness evaluation requires multiple structured summaries per query,
we will collect at least two (each with one or more sections) 
structured summaries per query by using multiple generative summarisers.
Moreover, from these structured summaries, we will automatically create \emph{degraded versions} of the summaries
using strategies such as the following:
\begin{description}
\item[NoHeadings] Remove all section headings from the original summary.
Comparing the original summary with a \textbf{NoHeadings} version in terms of SGSS
should be able to tell us the importance of the section headings.
Note that \textbf{NoHeadings} summaries will receive zero HR scores (Eq.~\ref{eq:HRep}).
\item[NoSection1] Remove the first section from the original summary.
As the first section of a structured summary is generally expected to be at least as important as the other sections,
if we compare the original summary with its \textbf{NoSection1} version, 
we expect SGSS to generally prefer the original over the \textbf{NoSection1} version
(unless the quality of the first section is low).
\end{description}

We will construct \emph{summary pair} data for a query set,
where the two structured summaries represent two different generative summarisers.
To have human labellers compare the two summaries side by side,
we will develop an annotation interface similar to that for comparing traditional ten-blue-links (e.g.~\cite[Figure 1]{sakai2021tois}).
We will assign $H (>2)$ human labellers to each summary pair,
and 
each human labeller will provide the following labels.
\begin{itemize}
\item OS (Overview Sufficiency), OF (Overview Faithfulness), and OR (Overview Representativeness) labels for the overviews of the two summaries (3-point scale);
\item An HR (Heading Representativeness) label for each heading of the two summaries (3-point scale);
\item SRel (Statement Relevance) and SF (Statement Faithfulness) labels for each statement of the two summaries (3-point scale);
\item An absolute Comprehensiveness label for each summary (3-point scale);
\item A \emph{preference label} which declares which of the two summaries is preferred (LEFT or RIGHT).
\end{itemize}
In addition, we will have LLMs provide the same set of labels.
We can then average the labels over the (human or combined) labellers.
In addition, we can \emph{derive} similar data for the \emph{degraded} summaries:
we assume that an original summary is preferred over its \textbf{NoHeadings} or \textbf{NoSection1} version (if the removed section contained at least one
relevant statement);
the other \emph{absolute labels} for the original summary mentioned above 
can simply be inherited to its degraded version.

The dataset thus constructed will be split into a training part and a test part: 
the former will be utilised for tuning purposes, i.e., for setting the weights,
and the latter will be used for evaluating and validating the SGSS framework.

Besides the above dataset, we will utilise a commercial search engine query log to explore 
methods for estimating the maximum time the user spends on a structured summary,
to determine $L_{\mathrm{max}}$ for computing XUX (See Section~\ref{ss:XUX}).

\subsection{Implementing SGSS}

It can be observed from the formulas presented in Section~\ref{s:proposed}
that SGSS
is a linear combination 
of several measures,
with weights $w_{\mathrm{OS}}, w_{\mathrm{OF}}, w_{\mathrm{OR}},
w_{\mathrm{HR}},
w_{\mathrm{SRel}}, w_{\mathrm{SF}},
w_{\mathrm{Comp}}$.
Hence,
if the SGSS scores of a given pair of summaries from the above dataset
are denoted by $\textit{SGSS}(S^{\mathrm{LEFT}})$ and $\textit{SGSS}(S^{\mathrm{RIGHT}})$,
the following is also a linear combination of the same measures with the above weights:
\begin{equation}\label{eq:DeltaSGSS}
\Delta \textit{SGSS} = \textit{SGSS}(S^{\mathrm{LEFT}}) - \textit{SGSS}(S^{\mathrm{RIGHT}}) \ .
\end{equation}
Since we have a preference label for each summary pair as well as the absolute labels 
for the overviews, headings, and statements in the summaries,
we can apply (say) logistic regression
to the training portion of the dataset to determine the weights,
where the target variable is the probability that
$S^{\mathrm{LEFT}}$ is better than $S^{\mathrm{RIGHT}}$.
This will give us our first instantiation of SGSS.

We will also consider model selection: can we simplify the definition of SGSS
by removing one or more evaluation criteria?

\subsection{Evaluating SGSS}

We would like SGSS to agree with human summary preferences.
To check this,
we can use the test portion of our dataset (which was not used for tuning the weights),
and compute the \emph{agreement rate} for each summary pair:
\begin{equation}\label{eq:AR}
\textit{AR} = \frac{
\text{\#human preference labels that agree with } \Delta \textit{SGSS}
}
{
H
} \ .
\end{equation}
Here,
although we have both human and LLM-based labels for the test data,
we will have to check whether relying only on LLM-based labels suffices,
because we would like to be able to 
\emph{automatically} estimate whether a given new summary is better than another.
(We will first have to examine the reliability of LLM-based labels 
by comparing them with the human labels.)
We will then take the mean of AR over the summary pairs in the test set to 
obtain the Mean AR (MAR)~\cite{sakai2021tois}.
We want the MAR of SGSS to be as high as possible.

%\subsection{Implementing the SGSS Framework}\label{ss:implementingSGSS}

%\subsection{Evaluating the SGSS Framework}\label{ss:evaluatingSGSS}

\section{Summary}\label{s:summary}

We proposed a framework
for evaluating structured generative search summaries
that are placed atop organic web search results, and described our plans for implementing and evaluating the framework.
The outcomes of our implementation and evaluation will be reported elsewhere.

% limitations

% single-turn SERPs

% Korean

% textual

%\subsection*{APPENDIX}

\section*{Acknowledgement}

This work was supported by Waseda University's Global Research Center (GRC)  Support Program (Project No. GRC-kojin-2504)
and JST SPRING (Grant Number JPMJSP2128).

%To be disclosed upon paper acceptance.
%We thank Dr. Young-In Song (Naver Corporation) for overseeing this industry-academia collaboration 

%and for his feedback on the paper.

%\section*{Ethical Considerations}

\clearpage
\bibliographystyle{ACM-Reference-Format}

\balance 
%\bibliography{sample-base}
\bibliography{arxiv2026sgss}

%\newpage

%\appendix

%This is a section in the appendix.

\end{document}